\documentstyle[pra,aps,multicol]{revtex}

\begin{document}

\title{Error Prevention Scheme with Four Particles}
\author{Lev Vaidman, Lior Goldenberg and Stephen Wiesner}
\address{School of Physics and Astronomy\\
Raymond and Beverly Sackler Faculty of Exact Sciences \\
Tel-Aviv University, Tel-Aviv 69978, Israel.}
\date{March 28, 1996}
\maketitle

\begin{abstract}
It is shown that a simplified version of the error correction code recently 
suggested by Shor exhibits manifestation of the quantum Zeno effect. Thus, 
under certain conditions, protection of an unknown quantum state is achieved. 
Error prevention procedures based on four-particle and two-particle encoding 
are proposed and it is argued that they have feasible practical 
implementations.
\end{abstract} 


\begin{multicols}{2}


Recently, Peter Shor \cite{Shor} obtained a surprising and even seemingly 
paradoxical result: an unknown quantum state can be kept unchanged in a noisy 
environment. His procedure was optimized by others \cite{CS,Steane,LMPZ,BS}, 
all trying to solve the problem of decoherence in quantum computers. However, 
at the present stage these results have more theoretical than practical 
importance, since they assume the existence of a fairly sophisticated quantum 
computer which has not been built yet. Shor's idea has another application 
which is more close to practical applications of today. Storing an unknown 
quantum state for some period of time (a ``quantum memory'') is an essential 
ingredient of various quantum cryptographic schemes \cite{Mor,GV}. It also can 
improve the reliability of the transmission of a quantum state \cite{EM}. The 
present technology is very close to the practical realization of these ideas 
\cite{Townsed,Kimble,Cirac95,Pellizzari95,Monroe}.

The reason why one can see Shor's result as paradoxical is the following. The 
expectation value of an operator can be measured using weak adiabatic 
measurements \cite{AV,AAnV}. Even if the coupling is very weak it can lead to 
a definite result, provided it is applied for a long enough time. This weak 
interaction with the measuring device can be considered as an action of a 
noisy environment, and therefore Shor's procedure performed frequently during 
the measurement should apparently keep the quantum state unchanged. Consider 
two eigenstates of an operator $A$ with the eigenvalues $a_1$, and $a_2$. If 
the initial state of the measured system is $|a_1\rangle$, then the outcome of 
the measurement will be $\langle A \rangle = a_1$; if, instead, the initial 
state is $|a_2\rangle$, then the outcome of the measurement will be 
$\langle A \rangle = a_2$; and if the initial state is $1/\sqrt2 \; 
(|a_1\rangle + |a_2\rangle)$, then the outcome of the measurement will be 
$\langle A\rangle = 1/2 \; (a_1 + a_2)$. However, a pointer showing 
$1/2 \; (a_1 + a_2)$ is a physical situation which is different from the 
mixture of situations in which the pointer shows $a_1$ and $a_2$. Therefore, 
we face a contradiction with the linearity of quantum theory.

The solution to this apparent paradox is that the coupling necessary for the 
adiabatic measurement of $A$, even if it is very weak, is different from the 
noise which can be dealt with using Shor's method. In the Shor procedure a 
qubit is encoded in nine particles, and the noise ``felt'' by each particle is 
assumed to be independent. However, in general, the variable $A$ is related to
several particles, such that the measurement requires coupling to some of them 
simultaneously. If we bring the particles to the same location and perform the 
adiabatic coupling with the measuring device, the independence requirement is 
not fulfilled explicitly. For some non-local variables there are measurement 
methods which can be applied without moving the particles to the same location 
\cite{AAV}, but then the parts of the measuring device which interact with the 
various particles must be in a correlated state prior to the interaction. 
Again, this corresponds to a correlated noise, and therefore Shor's procedure 
is not applicable.

There is another aspect of Shor's method which did not get enough attention. 
Usually, Shor's procedure and its modifications are considered as error {\it 
correction} schemes. Indeed, these methods correct the state completely if 
only one particle has decohered. But, Shor's procedure is also a scheme for 
{\em preventing} errors. For any system and any finite time $T$, frequent 
enough performance of Shor's procedure will ``freeze'' the state, exhibiting 
manifestation of the quantum Zeno effect. The use of the Zeno effect for 
correcting errors in quantum computers was first suggested by Zurek 
\cite{Zurek}, and it is a part of a scheme recently proposed by Barenco {\it 
at el.} \cite{BBDEJM}.

The only assumptions required for freezing a state are the boundness of the 
energy uncertainty of the system and the environment, and the independence of 
the noise disturbing each particle. If the energy uncertainty is bounded, then 
the rate of change of the quantum state of the system plus the environment is 
bounded. Given $N$, the number of Shor's tests performed during the time $T$, 
the evolution of a qubit and its environment $|e\rangle$ during a short period 
of time $T/N$ can be written in the following way:
\begin{eqnarray} 
|0\rangle |e\rangle & \longrightarrow & \gamma_1 |0\rangle |e\rangle 
+ \delta_1 |0\rangle |e_1\rangle + \delta_2 |1\rangle |e_2\rangle , \nonumber 
\\
|1\rangle |e\rangle & \longrightarrow & \gamma_2 |1\rangle |e\rangle 
+ \delta_3 |1\rangle |e_3\rangle + \delta_4 |0\rangle |e_4\rangle . 
\label{qubit+env-evol}
\end{eqnarray}
where $\langle e |e_1\rangle = \langle e |e_3\rangle = 0$, $\gamma_i = 1 - 
O(1/N^2)$ and $\delta_j = O(1/N)$. 
This evolution differs from what was considered before in that that no change 
takes place in the zero order in $1/N$. We consider the situation in which all 
the particles evolve (independently) according eq.(\ref{qubit+env-evol}) with 
different coefficients and different final states of their local environments. 
We claim that if Shor's tests are performed frequently enough they will always 
yield the outcome ``no error", and, in the same limit of large number of tests 
$N$, the state of the system will not be changed during the finite time $T$ 
(and this is what led us see the apparent paradox of Shor's scheme described 
above).

We propose to simplify Shor's procedure. We keep the encoding step and
the measurement step, but we suggest that we omit the correction step.
Moreover, the last part of the measurement step, the observation, can
be omitted too -- the coupling with the measuring device is enough.
Clearly, a protective device which only tests a system is
significantly simpler than one which also makes corrections. However,
the most important point is that we can reduce the number of particles
involved in the encoding of a quantum state. In the original Shor
procedure three triplets are used for determining which particle was
damaged: three particles or three triplets are compared, and if one is
different from the other two the state is corrected. In the present
procedure we do not have to perform corrections, so two doublets are
enough. Therefore, we need only four particles instead of nine. The
encoding is given by the following transformation
\begin{eqnarray} 
|0\rangle  & \longrightarrow &|0_{E}\rangle \equiv {1\over 2} \:
(|00\rangle + |11\rangle) \:(|00\rangle + |11\rangle) , \nonumber \\
|1\rangle  & \longrightarrow &|1_{E}\rangle \equiv {1 \over 2} \:
(|00\rangle - |11\rangle)\:(|00\rangle - |11\rangle) , \label{01-encoding}
\end{eqnarray}
such that a qubit $\alpha |0\rangle + \beta |1\rangle$ is encoded as  
$\alpha |0_{E}\rangle + \beta |1_{E}\rangle$.

The protection procedure consists of frequent tests that the four-particle 
system has not left the subspace of the encoded states given by 
eq.(\ref{01-encoding}). In order to see that an arbitrary encoded state 
$\alpha |0_{E}\rangle + \beta |1_{E}\rangle$ is indeed frozen due to the 
quantum Zeno effect we note that the two following conditions are fulfilled: 
First, after the evolution for a short time $T/N$ the amplitude of the state 
outside the subspace of the encoded states is of the order of $1/N$. The 
probability for getting the result ``out of subspace'' is of the order of 
$1/N^2$, and therefore, the probability of obtaining such an outcome during 
the time $T$ is proportional to $1/N$. Taking $N$, the number of tests during 
the time $T$, large enough we can decrease the probability of such an error 
below any desired level. Second, after the evolution of time $T/N$ the 
amplitude of the state $\beta^* |0_{E}\rangle - \alpha^* |1_{E}\rangle$, 
which is the state inside the subspace of the encoded states orthogonal to the 
initial encoded state, is of the order of $1/N^2$. Therefore, given that all 
$N$ projections yielded ``inside the subspace'', the difference between the 
final and the initial states is of the order of $1/N$ and can be neglected for 
large $N$.

The required projection can be performed in several steps which are, at least 
conceptually, simple. Each step is a certain {\em nonlocal} measurement in the 
sense that we measure a nonlocal variable related to two or more particles. We 
can do it without bringing the particles of the system together using 
correlated particles of the measuring device\cite{AV}, but in fact, since the 
simultaneity of the coupling with different particles of the system is not 
crucial for our purpose, we can use even single-particle measuring devices. 

The first step is to test that there are no terms which include the states 
$|01\rangle$ and $|10\rangle$ for the first two particles. The method is as 
follows: A test particle, prepared in a certain state, interacts with the 
first and then with the second particle of the system. The interaction 
``flips'' the state of the test particle if the particle of the system is in 
the state $|1\rangle$ and does not flip it if the particle is in the state 
$|0\rangle$. The states $|0\rangle$ and $|1\rangle$ of the particles of the 
system remain unchanged by this procedure. Then we measure the final state of 
the test particle. If this state is identical to the initial state we know 
that the system has only terms of the form $|00\rangle$ and $|11\rangle$. If 
we perform our tests frequently enough the probability to find ``wrong'' terms 
during all the tests goes to zero. In this case we can omit the last part 
of the procedure since the quantum Zeno effect requires only correlation 
with some external system, and therefore, the observation of the state of the 
test particle is not necessary. It is interesting to note that it is also not 
necessary to prepare a well defined initial state of the test particle. 
Although there are certain initial states of the test particle which end up 
uncorrelated to the system, several test particles emerging from a truly 
random source will work too.  

The next step is the same procedure performed with particles 3 and 4. After 
completing this test we know that the state of the four-particle system has 
the form
\begin{eqnarray} 
& &  a (|00\rangle + |11\rangle) \:( |00\rangle +  |11\rangle) \nonumber \\
& + & b (|00\rangle - |11\rangle) \:( |00\rangle -  |11\rangle) \nonumber \\ 
& + & c (|00\rangle + |11\rangle) \:( |00\rangle -  |11\rangle) \nonumber \\
& + & d (|00\rangle - |11\rangle) \:( |00\rangle +  |11\rangle) . \label{2}
\end{eqnarray}
For completing the projection on the subspace of the encoded states in 
eq.(\ref{01-encoding}) we have to show that the coefficients $c$ and $d$ 
vanish. In order to see how this can be achieved we rewrite the state in 
eq.(\ref{2}) using new local bases for all particles, $|\tilde 0\rangle
\equiv 1/\sqrt2 \: (|0\rangle + |1\rangle)$ and $|\tilde 1\rangle
\equiv1/\sqrt 2 \: (|0\rangle - |1\rangle)$,
\begin{eqnarray} 
& &   a (|\tilde 0\tilde 0\rangle + |\tilde 1\tilde 1\rangle) \: 
        (|\tilde 0\tilde 0\rangle + |\tilde 1\tilde 1\rangle) \nonumber \\
& + & b (|\tilde 0\tilde 1\rangle + |\tilde 1\tilde 0\rangle) \: 
        (|\tilde 0\tilde 1\rangle + |\tilde 1\tilde 0\rangle) \nonumber \\
& + & c (|\tilde 0\tilde 0\rangle + |\tilde 1\tilde 1\rangle) \: 
        (|\tilde 0\tilde 1\rangle + |\tilde 1\tilde 0\rangle) \nonumber \\
& + & d (|\tilde 0\tilde 1\rangle + |\tilde 1\tilde 0\rangle) \: 
        (|\tilde 0\tilde 0\rangle + |\tilde 1\tilde 1\rangle). \label{3}
\end{eqnarray}
The final step of the procedure is similar to the first two steps, but it 
involves interaction with all four particles of the system. The test particle, 
prepared in a certain state, interacts with all four particles one after the 
other in such a way that it ``flips'' if the particle is in the state 
$|\tilde 1\rangle$ and does not ``flip'' if the particle is in the state 
$|\tilde 0\rangle$. We ``look'' on the test particle only after it has 
interacted with all four particles. If the final state of the test particle is 
identical to the initial one, we know that the state of the system belongs to 
the desired subspace. Again, observing the test particle at the end of the
process is not necessary for the Zeno effect to occur. 

The Shor error correcting method was optimized such that only five particles 
instead of nine are used \cite{LMPZ,BS}. Thus, one might expect that the ideas 
of five-particle encoding procedures can be used to reduce the number of 
particles necessary for the Zeno-type error prevention method. However, the 
simple counting argument used to show that five is the minimal number of 
particles required for error correction \cite{LMPZ,EM} suggests that three 
qubits are not enough for error prevention. A necessary requirement for an 
error prevention code is that at the first order, i.e. via one-particle 
decoherence, the state $|0_{E}\rangle$ cannot evolve to the state 
$|1_{E}\rangle$. If we assume that each type of decoherence (flip, sign 
change, and flip together with sign change) moves an encoded ``0'' to 
different orthogonal states, then they are $2^3 - 1 -3 \times 3 = -2$ 
available states for the encoded ``1''. This, of course, is a meaningless 
statement since the assumption of the orthogonality of the states created 
after various one-particle decoherence actions is wrong. Nevertheless, the 
general statement remains true and can be checked by a straightforward 
analysis: all possible three-particle encodings of one state generate, via
single-particle decoherence, enough states to cover the whole Hilbert space of 
the three-particle system. Note that for four-particle encoding this naive 
counting shows no problem; we have $2^4 - 1 - 4 \times 3 = 3$ orthogonal 
states for encoding ``1''.

In fact, some single-particle errors create identical vectors such that we 
have only six mutually orthogonal states obtained from the state 
$|0_E\rangle$. Moreover, one error yield only four new states from the state 
$|1_E\rangle$. So we still have $2^4 - 1 - 6 - 1 - 4 = 4$ states available. 
As pointed out by Shor \cite{Shor-pc}, this allows us to encode one more 
qubit using the same four particles. The states of a system composed by two 
qubits can be encoded as follows:
\begin{eqnarray}
|00\rangle  & \longrightarrow & |0_{E}\rangle \equiv {1\over 2} \:
(|00\rangle + |11\rangle) \: (|00\rangle + |11\rangle) , \nonumber \\
|01\rangle  & \longrightarrow & |1_{E}\rangle \equiv {1\over 2} \:
(|00\rangle - |11\rangle) \: (|00\rangle - |11\rangle) , \nonumber \\
|10\rangle  & \longrightarrow & |2_{E}\rangle \equiv {1\over 2} \:
(|01\rangle + |10\rangle) \: (|01\rangle + |10\rangle) , \nonumber \\
|11\rangle  & \longrightarrow & |3_{E}\rangle \equiv {1\over 2} \:
(|01\rangle - |10\rangle) \: (|01\rangle - |10\rangle) . \label{04-encoding}
\end{eqnarray}
There is no single particle error which can bring from one encoded state 
$|i_E\rangle$ ($i = 0,1,2,3$) to another. Thus, frequent projections on the
subspace generated by the four states $|i_E\rangle$ should protect an unknown 
state in four-dimensional Hilbert space, i.e. two qubits. The realization of 
this projection is only slightly more difficult than the projection on the 
two-state subspace. The last step (the four-particle measurement) remains the 
same, but instead of the two two-particle tests we need to perform one 
four-particle measurement similar to that of the second step, but in the 
original basis.

We have shown that encoding one or two qubits in four qubits is in principle 
enough for the error prevention procedure. However, it is important to examine 
the type of noise in our system. Our method relies on the Zeno effect so it 
can deal only with ``slow" noise. The characteristic time of the noise 
coupling has to be larger than the time interval between the projection 
measurements. If the realistic model of the noise is that molecules of the 
environment cause very fast and large uncertain changes during rare collisions 
with the particles, then our method is not applicable. It also cannot help if
the main cause of the decoherence is some spontaneous decay process, since the 
quantum Zeno effect does not take place when the time interval between the 
measurements is larger than the characteristic time for which the exponential 
decay approximation is applicable. However, if the appropriate model is that 
the environment becomes slowly entangled with the system, then our method 
works.

Even in this case the error correction codes have some advantages over error 
prevention codes. The frequency of the required procedures is significantly 
smaller in the error correction code; the error correction procedure has to be
performed before the time that the second order disturbance becomes large, 
while the error prevention procedure has to be performed before the time that 
the first order disturbance becomes large. Thus, if we want to slow-down the 
decoherence by a factor of $N$, we have to perform our error-prevention 
procedure by the same factor $N$ more frequently than an error correction 
code. However, since our procedure is much more simple, it is very plausible 
that it will be more practical in some cases. In particular, since the 
technology of handling several qubits is just developing, it is most probable 
that the first experiments will be performed with a minimal number of 
entangled particles.

There are quantum systems for which the noise leads mainly to dephasing, 
leaving the amplitude unchanged. This happens when the orthogonal states in
a {\it particular} basis become entangled with the environment. It has been 
shown \cite{EM,Sam} that for this restricted type of decoherence there exist 
three-particle error correction codes. The quantum Zeno effect can help in 
this case too, and we suggest a two-particle error prevention scheme. The 
encoding is given by    
\begin{eqnarray} 
|0\rangle  & \longrightarrow & |0_{E}\rangle \equiv  1/\sqrt2 \: 
(|00\rangle + |11\rangle) , \nonumber \\
|1\rangle  & \longrightarrow & |1_{E}\rangle \equiv 1/\sqrt2  \: 
(|01\rangle + |10\rangle) . \label{02-encoding}
\end{eqnarray}
The error prevention procedure is especially simple in this case. We have to 
test that our state belongs to the subspace generated by the two encoded 
states $|0_{E}\rangle$ and $|1_{E}\rangle$. This can be implemented using just 
a single step of the type described above. Again, the particle of the 
measuring device is prepared in a certain state, then it interacts with the 
two particles of the system, one after the other. The interaction is such that 
the state of the test particle flips if the particle is in the state 
$|\tilde 1\rangle$ and does not flip if the particle is in the state 
$|\tilde 0\rangle$. If the state of the system belongs to the subspace of 
the encoded states, which can be written in the form $ a|\tilde 0\tilde 
0\rangle + b |\tilde 1\tilde 1\rangle$, the state of the test particle will 
not be flipped after the two interactions, while it will be flipped if the 
state of the two particles does not belong to this subspace. This correlation 
leads to the quantum Zeno effect which effectively prevents the system 
leaving the subspace of encoded states. Moreover, the state does not change 
significantly inside the subspace. The phase error after one such operation is 
of the order of $1/N^2$, and therefore, for a large number of tests $N$ during 
the period of time $T$ the total error can be neglected.

Maybe in a somewhat pessimistic tone we want to conclude saying that the real
problem with error preventing or correcting codes is the noise introduced by 
the procedure itself. As was explained at the beginning, the type of 
interaction involved in the prevention-correction measurements requires either 
bringing the particles of the system together, or letting them interact with 
correlated particles or with a single particle as proposed here. In all these 
cases the noise, if present, cannot be considered independent, and therefore 
the error correction or prevention effects of all the discussed methods do not 
occur. This does not mean that the result of Shor is not important -- even the 
reduction of decoherence between the measurements is an extremely important 
and surprising effect.

Taking into account the price in the noise which we will probably have to pay 
in every projection procedure, the fact that we have to perform them more 
frequently is a significant disadvantage of error prevention schemes over 
error correction schemes. But again, it is compensated for by the fact that we 
need a smaller number of steps for each projection procedure. Since our code 
seems to be the simplest code proposed so far, it has a good chance to be the 
first implemented in a real laboratory. The experimental observation of the 
quantum Zeno effect reported by Itano {\it et al.} \cite{Itano} contributes to 
our optimism. Thus, our scheme might serve as an effective testbed for the 
robustness of quantum computers and other quantum communication devices.

This research was supported in part by grant 614/95 of the Basic Research 
Foundation (administered by the Israel Academy of Sciences and Humanities).



\end{multicols}

\end{document}